\documentclass[aps,a4paper,superscriptaddress,preprintnumbers,showpacs,
amsmath,amssymb]{revtex4}
\usepackage{graphicx}
\usepackage{dcolumn}
\usepackage{color}
\usepackage{latexsym,amsfonts}
\usepackage{textcomp}
\usepackage{bm}

\usepackage{ulem}

\begin{document}

\title{Probing of violation of Lorentz invariance\\ by ultracold
  neutrons in the Standard Model Extension}

\author{A. N. Ivanov}\email{ivanov@kph.tuwien.ac.at}
\affiliation{Atominstitut, Technische Universit\"at Wien, Stadionallee
  2, A-1020 Wien, Austria}
\author{M. Wellenzohn}\email{max.wellenzohn@gmail.com}
\affiliation{Atominstitut, Technische Universit\"at Wien, Stadionallee
  2, A-1020 Wien, Austria} \affiliation{FH Campus Wien, University of
  Applied Sciences, Favoritenstra\ss e 226, 1100 Wien, Austria}
\author{H. Abele}\email{abele@ati.ac.at}
\affiliation{Atominstitut, Technische Universit\"at Wien, Stadionallee
  2, A-1020 Wien, Austria}

\date{\today}

\begin{abstract}
We analyze a dynamics of ultracold neutrons (UCNs) caused by
interactions violating Lorentz invariance within the Standard Model
Extension (SME) (Colladay and Kosteleck\'y, Phys. Rev. D {\bf 55},
6760 (1997) and Kosteleck\'y, Phys. Rev. D {\bf 69}, 105009
(2004)). We use the effective non--relativistic potential for
interactions violating Lorentz invariance derived by Kosteleck\'y and
Lane (J. Math. Phys. {\bf 40}, 6245 (1999)) and calculate
contributions of these interactions to the transition frequencies of
transitions between quantum gravitational states of UCNs bouncing in
the gravitational field of the Earth. Using the experimental
sensitivity of qBounce experiments we make some estimates of upper
bounds of parameters of Lorentz invariance violation in the neutron
sector of the SME which can serve as a theoretical basis for an
experimental analysis. We show that an experimental analysis of
transition frequencies of transitions between quantum gravitational
states of unpolarized and polarized UCNs should allow to place some
new constraints in comparison to the results adduced by Kosteleck\'y
and Russell in Rev. Mod. Phys. {\bf 83}, 11 (2011); edition 2019,
arXiv: 0801.0287v12 [hep-ph].
\end{abstract}
\pacs{11.10.Ef, 11.30.Cp, 12.60.-i, 14.20.Dh}

\maketitle

\section{Introduction}
\label{sec:introduction}

As has been pointed out by Kosteleck\'y \cite{Kostelecky2004}, the
combination of Einstein's general relativity or a theory of
gravitation and the Standard Model (SM) of particle physics provides a
remarkably successful description of nature. In such a combined theory
gravitation and dynamics of SM particles are described at the
classical and quantum level, respectively. It may be expected that
these two theories can be merged into a unified quantum theory at the
Planck scale $m_{\rm Pl} = 1/\sqrt{G_N} \sim 10^{19}\,{\rm GeV}$,
where $G_N$ is the Newtonian gravitational constant \cite{PDG2018},
and the effects, which might be originated at the Planck scale, of
such a unified quantum field theory may be associated with the
breaking of Lorentz symmetry and be observable at low--energy scales,
described using effective quantum field theory called the Standard
Model Extension (SME) \cite{Kostelecky1997a, Kostelecky1998a,
  Kostelecky2006, Kostelecky2011a}. At the classical level, the
dominant terms in the SME include the pure--gravity and minimally
coupled SM action, together with all leading--order terms introducing
violations of Lorentz symmetry that can be contracted from
gravitational and SM fields. According to L\"uders--Pauli theorem
\cite{Streater1980} (see also \cite{Greenberg2002}) violation of
Lorentz invariance entails violation of CPT invariance, where $C$, $P$
and $T$ are charged conjugation, parity and time reversal
transformations, respectively.

For experimental searches of violation of Lorentz invariance there has
been proposed i) to compare anomalous magnetic moments of the electron
and positron \cite{Kostelecky1997b}, ii) to use the Penning-trap
experiments, measuring anomaly frequencies and providing the sharpest
tests of CPT symmetry \cite{Kostelecky1998b}, iii) to compare hydrogen
and antihydrogen spectroscopy \cite{Kostelecky1999a, Kostelecky2015},
iv) to use existing data on the ground state of muonium and on the
muon anomalous magnetic moment \cite{Kostelecky2000, Kostelecky2014},
and v) to measure neutrino spectra in beta decays
\cite{Kostelecky2013}. Contemporary experimental data on violation of
Lorentz invariance are gathered in \cite{Kostelecky2011b}.

The use of UCNs, quantized in the gravitational field of the Earth
\cite{Nesvizhevsky2000, Nesvizhevsky2002, Nesvizhevsky2003} (see also
\cite{Nesvizhevsky2005}), was proposed for experimental searches of
gravitational effects such as limits on non-Newtonian interactions in
the range $1 - 10 \,{\rm microns}$ by Abele {\it et al.}
\cite{Abele2003} and by Nesvizhevsky and Protasov
\cite{Nesvizhevsky2004} and weak equivalence principle by Bertolami
and Nunes \cite{Bertolami2003}, and by Pokotilovski
\cite{Pokotilovski2012}. Then, the method of the quantum gravitational
states of UCNs bouncing in the gravitational field of the Earth
\cite{Abele2009, Jenke2009}, named as Gravity Resonance Spectroscopy
and based on the method of Resonance Spectroscopy or the ``molecular
beam resonance method'' introduced by Rabi {\it et al.}
\cite{Rabi1939}, has been applied in \cite{Abele2010} --
\cite{Cronenberg2018} to experimental searches of large variety of
gravitational effects.

Recently Mart\'in-Ruiz and Escobar \cite{Escobar2018, Escobar2019}
have used UCNs, quantized in the gravitational field of the Earth, as
a tool to probe effects of violation of Lorentz invariance using the
experimental data on the energy levels of the ground $E_1$ and first
excited $E_2$ quantum gravitational states of UCNs
\cite{Nesvizhevsky2005}.  An influence of CPT--violating effects on
the quantum gravitational states of UCNs has been also investigated by
Zhi Xiao \cite{Xiao2019}.  The analysis, carried out by Mart\'in-Ruiz
and Escobar \cite{Escobar2018, Escobar2019}, has allowed to impose the
following upper bounds on the parameters of violation of Lorentz
invariance
\begin{eqnarray}\label{eq:1}
  3\,\frac{\Delta E^{\exp}_k}{E^{\exp}_k} > |\bar{c}^n_{00} - 2
  \bar{c}^n_{zz}| \quad \textrm{\cite{Escobar2018}},\quad
  3\,\frac{\Delta E^{\exp}_k}{E^{\exp}_k} > |3\bar{s}^n_{00} +
  \bar{s}^n_{zz}| \quad ~\textrm{\cite{Escobar2019}},
\end{eqnarray}
where $E^{\exp}_k$ are the experimental values of the energy levels of
UCNS in the quantum gravitational ground $k = 1$ and first excited $k
= 2$ states, respectively, and $\Delta E^{\exp}_k$ are their
sensitivities. The parameters $c^n_{\mu\nu}$ and $s^n_{\mu\nu}$ define
the strength of violation of Lorentz invariance in the kinetic term of
the neutron and the kinetic term of the gravitational field in the
Einstein--Hilbert action, respectively.  In the Dirac action for the
neutron field in a weak gravitational field such as the gravitational
field of the Earth the $\gamma^{\mu}$--matrix is replaced by
$\gamma_{\mu} \to \gamma_{\mu} + c^n_{\nu\mu}\gamma^{\nu} + \ldots$
with a neglect of contributions of a weak gravitational field to the
terms violating Lorentz invariance \cite{Kostelecky1997a,
  Kostelecky1998a}, where the ellipsis stands for another
contributions violating Lorentz invariance.  The index ``n'' implies
that such a coefficient $c^n_{\mu\nu}$ violating Lorentz invariance
can be observable only in experiments with neutrons. In turn, the term
$s^{n \mu\nu}R^T_{\mu\nu}$ enters to the effective action of the
minimal gravity SME (with vanishing torsion) in the form $R \to R +
s^{n \mu\nu} R^T_{\mu\nu}$ \cite{Kostelecky2004, Kostelecky2006,
  Kostelecky2011a}, where $R$ and $R^T_{\mu\nu}$ are a scalar
curvature and a traceless Ricci tensor, respectively.  Then, the
coefficients $c^n_{\mu\nu}$ and $s_{\mu\nu}$ are defined as
$c^n_{\mu\nu} = \bar{c}^n_{\mu\nu} + \tilde{c}^n_{\mu\nu}$ and
$s^n_{\mu\nu} = \bar{s}^n_{\mu\nu} + \tilde{s}^n_{\mu\nu}$
\cite{Kostelecky2004, Kostelecky2006, Kostelecky2011a}, where
$\bar{c}^n_{\mu\nu}$ and $\bar{s}^n_{\mu\nu}$ are the vacuum
expectation values, whereas $\tilde{c}^n_{\mu\nu}$ and
$\tilde{s}^n_{\mu\nu}$ define fluctuations around vacuum expectation
values \cite{Kostelecky2004, Kostelecky2006,
  Kostelecky2011a}. Of course, the fluctuations can in
  principle contribute to experimental effects \cite{ Kostelecky2006,
    Kostelecky2011a}. However, in our work we neglect the
  contributions of fluctuations in comparison to contributions of
  vacuum terms \cite{Kostelecky2006}.

This paper is addressed to extraction of an information on violation
of Lorentz invariance from the analysis of experimental data on
transition frequencies of transitions between quantum gravitational
states of unpolarized UCNs, obtained in \cite{Cronenberg2018}. Using
the current sensitivity $\Delta E < 2\times 10^{-15}\,{\rm eV}$ of the
qBounce experiments, closely related to experimental uncertainties of
experimental data \cite{Cronenberg2018} (see a discussion below
Eq.(\ref{eq:16})), we place some constraints on parameters of Lorentz
invariance violation from corrections to transition frequencies of
non--spin--flip and spin--flip transitions between quantum
gravitational states of polarized UCNs. We analyze the contributions
  of interactions violating Lorentz invariance at the neglect of the
  contributions of the chameleon--neutron interactions
  \cite{Ivanov2013, Ivanov2015a, Ivanov2016, Jaffe2017} and
  symmetron--neutron interactions \cite{Cronenberg2018}, where scalar
  chameleon and symmetron fields were introduced in \cite{Khoury2004}
  and \cite{Khoury2010} as candidates for explanation of dynamics of
  the Universe such as an origin of dark energy, inflation and
  late--time acceleration.  Taking into account the constraints on the
  parameters $\bar{s}^n_{\mu\nu}$ adduced in
  Ref.\cite{Kostelecky2011b} (see p. 90, Table D40, Gravity sector, $d
  = 4$ (part 2 of 3)) we neglect the contributions of Lorentz
  invariance violation in the gravitational sector and analyze the
  effects of Lorentz invariance violation in the neutron sector
  only. For this aim we use the effective non--relativistic potential
  of interactions violating Lorentz invariance which has been derived
  by Kosteleck\'y and Lane \cite{Kostelecky1999b}.

The paper is organized as follows. In section \ref{sec:potential} we
discuss the general form of the relativistic Lagrangian for a free
neutron in the SME \cite{Kostelecky1997a, Kostelecky1998a}, and the
non--relativistic potential, derived by Kosteleck\'y and Lane
\cite{Kostelecky1999b}. In section \ref{sec:earth} we define a
location of the Institut Laue Langevin (ILL) in Grenoble on the
surface of the Earth, and calculate the corrections to the transition
frequencies of transitions between quantum gravitational states of
unpolarized and polarized UCNs in the standard laboratory frame.We
distinguish corrections to the binding energies of quantum
gravitational states of unpolarized and polarized UCNs. This is
because of the 2--fold degeneracy of the energy levels of quantum
gravitational states of unpolarized UCNs caused by spin--degrees of
freedom \cite{LL1965, Davydov1965}. In section \ref{sec:sun} we define
the parameters of Lorentz invariance violation in the canonical
Sun--centered frame. We use the current experimental sensitivity
$\Delta E < 2\times 10^{-15}\,{\rm eV}$ of the qBounce experiments and
impose constraints on parameters of Lorentz invariance violation
defined in the canonical Sun-centered frame extracted from the
corrections to the transition frequencies of transitions between
quantum gravitational states of unpolarized and polarized UCNS. In
section \ref{sec:spin} we derive Heisenberg's equation of a neutron
spin evolution, caused by interactions violating Lorentz
invariance. In section \ref{sec:discussion} we discuss the obtained
results and perspectives of the analysis of parameters of Lorentz
invariance violation in the qBounce experiments with an improved
sensitivity $\Delta E$.

\vspace{-0.15in}

\section{Effective non--relativistic potential for 
  Lorentz invariance violation in the neutron sector of the SME}
\label{sec:potential}

The general relativistic Lagrangian for a free neutron in the SME
takes the form \cite{Kostelecky1997a, Kostelecky1998a}
\begin{eqnarray}\label{eq:2}
 {\cal L}_{\rm SME} = \frac{1}{2}\,\bar{\psi}i\,\Gamma_{\nu}\!
 \stackrel{\leftrightarrow}{\partial^{\nu}}\psi - \bar{\psi} M
 \psi,
\end{eqnarray}

\noindent where $\Gamma_{\nu}$ and $M$ are given by \cite{Kostelecky1999b}
\begin{eqnarray}\label{eq:3}
\hspace{-0.3in} \Gamma_{\nu} = \gamma_{\nu} + c^n_{\mu\nu}
\gamma^{\mu} + d^n_{\mu\nu}\gamma^5 \gamma^{\mu} + e^n_{\nu} +
if^n_{\nu} \gamma^5 + \frac{1}{2}\,g^n_{\lambda \mu \nu}
\sigma^{\lambda \mu} \quad,\quad M = m + a^n_{\mu} \gamma^{\mu} +
b^n_{\mu} \gamma^5 \gamma^{\mu} + \frac{1}{2}\,H^n_{ \mu \nu}
\sigma^{\mu \nu}
\end{eqnarray}
with usual definition of the Dirac matrices $\{1, \gamma^{\mu},
\gamma^5, \gamma^5\gamma^{\mu}, \sigma^{\mu\nu}\}$ and the Minkowski
metric tensor $\eta^{\mu\nu}$ with a signature $(1, - 1,- 1,- 1)$
\cite{Itzykson1980}, and $m$ is a neutron mass \cite{PDG2018}.  The
parameters $a^n_{\mu}, b^n_{\mu}, c^n_{\mu\nu}, d^n_{\mu\nu},
e^n_{\mu}, g^n_{\lambda \mu \nu}$ and $H^n_{ \mu \nu}$ are responsible
for violation of Lorentz invariance. In an inertial frame of an
observer they can be treated as fixed real Lorentz vectors and tensors
\cite{Kostelecky1999b}. The tensors $H^n_{\mu\nu}$, $c^n_{\mu\nu}$ and
$d^n_{\mu\nu}$, and $g^n_{\lambda \mu \nu}$ are antisymmetric,
traceless, and antisymmetric with respect to first two indices
\cite{Kostelecky1999b}, respectively.

The non--relativistic potential $\Phi_{\rm n LV}$, describing effects
of violation of Lorentz invariance in the neutron sector of the SME,
has been derived by Kosteleck\'y and Lane \cite{Kostelecky1999b} by
using Foldy--Wouthuysen (FW) canonical transformations
\cite{Foldy1950} from the relativistic Lagrangian Eq.(\ref{eq:2}) to
order $O(|\vec{p}\,|^3/m^3)$, where $\vec{p}$ is a 3--momentum
operator of the neutron. It takes the form (see Eq.(26) of
Ref. \cite{Kostelecky1999b})
\begin{eqnarray}\label{eq:4}
\hspace{-0.15in}&&\Phi_{\rm nLV} = 2\,\Big(- b^n_{\ell} + m d^n_{\ell
  0} - \frac{1}{2}\,m \,\varepsilon_{\ell k j} g^n_{k j 0} +
\frac{1}{2}\,\varepsilon_{\ell k j} H^n_{k j}\Big)S_{\ell} + \big(-
a^n_j + m(c^n_{0j} + c^n_{j0}) + m e^n_j\big) \frac{p_j}{m} + 2
\,\Big(b^n_0 \delta_{j\ell} -m(d^n_{\ell
  j}\nonumber\\\hspace{-0.15in}&& + d^n_{00} \delta_{\ell j}) -
\frac{1}{2}\,m\, \varepsilon_{\ell k m}\big(g^n_{m k j} + 2
g^n_{m00}\delta_{j k}\big) - \varepsilon_{j \ell k} H^n_{k
  0}\Big)\frac{p_j}{m}S_{\ell} - \big(2 c^n_{jk} +
c^n_{00}\delta_{jk}\big)\frac{p_j p_k}{2 m} + \Big(\big(4 d^n_{0j} + 2
d^n_{j0} - \varepsilon_{j m n} g^n_{m n 0}\big)\,
\delta_{k\ell}\nonumber\\\hspace{-0.15in}&& + \varepsilon_{\ell m n}
g^n_{mn0}\, \delta_{jk} - 2\,\varepsilon_{j \ell m}\,\big(g^n_{m 0 k}
+ g^n_{m k 0}\big)\Big)\frac{p_j p_k}{2 m}\,S_{\ell} + \Big(\big( -
b_j - \frac{1}{2}\,\varepsilon_{j m n} H_{mn}\big)\,\delta_{k\ell} +
b_{\ell}\Big)\,\frac{p_j p_k}{m^2}\,S_{\ell} + \frac{1}{2}\,(a_j
\delta_{k\ell} - m e_j
\delta_{k\ell})\nonumber\\\hspace{-0.15in}&&\times\,\frac{p_j p_k
  p_{\ell}}{m^3} + \Big( \big(- b_0\delta_{j \ell} + m d_{\ell j} +
\varepsilon_{j \ell n }\,H_{n0}\big)\,\delta_{k m} + \big( - m d_{jk}
- m \frac{1}{2}\,\varepsilon_{k n p}\,g_{n p j}\big)\,\delta_{m
  \ell}\Big)\,\frac{p_j p_k p_m}{m^3}\,S_{\ell} + \ldots\,,
\end{eqnarray}
where $p_j = - i\nabla_j$ and $S_j = \frac{1}{2}\,\sigma_j$ are the
neutron 3--momentum and spin operators, respectively, and $\sigma_j$
are $2\times 2$ Pauli metrices \cite{Itzykson1980}. The
non--relativistic potential Eq.(\ref{eq:4}) is obtained in agreement
with general assumption that dominant contributions to effects of
violation of Lorentz invariance are linear in parameters of violation
of such an invariance. The ellipsis denotes the
  contribution of the terms which have been neglected by Kosteleck\'y
  and Lane \cite{Kostelecky1999b} for the derivation of the
  non--relativistic potential to order $O(|\vec{p}\,|^3/m^3)$.

The evolution of UCNs in such a theory is described by the
Schr\"odinger--Pauli equation
\begin{eqnarray}\label{eq:5}
 i\frac{\partial \Psi_{k\sigma}}{\partial t} = {\rm H}
 \Psi_{k\sigma}\quad, \quad {\rm H} = {\rm H}_0 + \Phi_{\rm nLV} =
 -\frac{1}{2 m}\,\Delta + m g z + \Phi_{\rm nLV}
\end{eqnarray}
where $\Psi_{k\sigma}$ is a two--component spinorial wave function of
UCNs in the $k$--gravitational state and in a spin eigenstate $\sigma
= \uparrow$ or $\downarrow$, $\Delta$ is the Laplacian operator, and
$m g z$ is the gravitational potential of the Earth with the standard
gravitational acceleration $g$ having the local Grenoble value $g =​
9.80507(2)\,{\rm m/s^2}$ \cite{Cronenberg2018}.

\section{Parameters of violation of Lorentz  invariance in the
 standard laboratory frame}
\label{sec:earth}

The experiments with UCNs, bouncing in the gravitational field of the
Earth, are being performed in the laboratory at Institut Laue Langevin
(ILL) in Grenoble. The ILL laboratory is fixed to the surface of the
Earth in the northern hemisphere. Following \cite{Kostelecky2002a,
  Kostelecky2002b, Kostelecky2003, Kostelecky2009, Kostelecky2016} we
choose the ILL laboratory or the standard laboratory
  frame with coordinates $(t, x, y, z)$, where the $x$, $y$ and $z$
axes point south, east and vertically upwards, respectively, with
northern and southern poles on the axis of the Earth's rotation with
the Earth's sidereal frequency $\Omega_{\oplus} = 2\pi/(23\,{\rm hr}\,
56\,{\rm min}\, 4.09\,{\rm s} = 7.2921159 \times 10^{-5}\,{\rm
  rad/s}$. The position of the ILL laboratory on the surface of the
Earth is determined by the angles $\chi$ and $\phi$, where $\chi =
90^0 - \theta$ is the colatitude of the laboratory, defined in terms
of the latitude $\theta$, and $\phi$ is the longitude of the
laboratory measured east of south with the values $\theta =
45.16667^0$\,N and $\phi = 5.71667^0$\,E \cite{Grenoble},
respectively. The beam of UCNs moves from south to north antiparallel
to the $x$--direction and with energies of UCNs quantized in the
$z$--direction. In this section we neglect the Earth's rotation
assuming that the laboratory frame is an inertial one. In Fig.\,1 we
show a location of the ILL laboratory on the surface of the Earth.

The transition frequency $\nu_{p\sigma'q\sigma}$ of the transition
between two gravitational states of polarized UCNs $|q\sigma\rangle
\to |p\sigma'\rangle$ is defined by $\nu_{p\sigma'q\sigma} =
(E_{p\sigma'} - E_{q\sigma})/2\pi$, where $E_{q\sigma}$ and
$E_{p\sigma'}$ are observable binding energies of UCNs in the
$|q\sigma\rangle$ and $|p\sigma'\rangle$ quantum gravitational states
with definite spin eigenstates. Contributions of interactions
violating Lorentz invariance to transition frequencies of transitions
$|q\sigma\rangle \to |p\sigma'\rangle$ between quantum gravitational
states $|q\sigma \rangle$ and $|p\sigma' \rangle$ of UCNs are defined
by corrections to the binding energies of these bound states. For
practical analysis we have to distinguish corrections to the binding
energies of unpolarized and polarized UCNs, respectively \cite{LL1965,
  Davydov1965}.

The measurement of transition frequencies of transitions $|1\rangle
\to |3\rangle$ and $|1\rangle \to |4\rangle$ between quantum
gravitational states $|1\rangle$, $|3\rangle$ and $|4\rangle$ of
unpolarized UCNs with experimental values $\nu^{\exp}_{31} = (E_3 -
E_1)/2\pi = 464.8(1.3)\,{\rm Hz} = 0.3059(8)\,{\rm peV}$ and
$\nu^{\exp}_{41} = (E_4 - E_1)/2\pi = 649.8(1.8)\,{\rm Hz} =
0.4277(12)\,{\rm peV}$, where $1\,{\rm peV} = 10^{-12}\,{\rm eV}$
\cite{PDG2018}, has been reported by Cronenberg {\it et al.}
\cite{Cronenberg2018}.  The $E_1$, $E_2$ and $E_4$ are observable
binding energies of quantum gravitational states of unpolarized UCNs
in the ground $|1\rangle$ and two excited $|3\rangle$ and $|4\rangle$
states, respectively.  Theoretical values of binding energies of
unperturbed quantum gravitational states of UCNs are defined by
$E^{(0)}_k = E_0|\xi_k|$ for the principal quantum number $k = 1,2,
\ldots$, where $\xi_k$ is the root of the wave function
$\psi^{(0)}_k(0) = 0$ \cite{Gibbs1975, Westphal2007}, which obeys the
Schr\"odinger equation $({\rm H}_0 - E^{(0)}_k)\,\psi^{(0)}_k(z) = 0$,
and $E_0 = m g\ell_0 = \sqrt[3]{m g^2/2} = 0.6016\,{\rm peV}$ is the
quantum scale of quantum gravitational states of UCNs \cite{Gibbs1975}
calculated for $m = 939.5654\,{\rm MeV}$ and $g = 9.80507\,{\rm
  m/s^2}$.  For the quantum gravitational states $|1\rangle$,
$|3\rangle$ and $|4\rangle$ we get $E^{(0)}_1 = 1.4066\,{\rm peV}$,
$E^{(0)}_3 = 3.3212\,{\rm peV}$ and $E^{(0)}_4 = 4.0829\,{\rm
  peV}$. The experimental values of transition frequencies are
measured with relative uncertainties $2.6\times 10^{-3}$ and
$2.8\times 10^{-3}$, respectively.  Since $\xi_1$, $\xi_3$ and $\xi_4$
are well--defined mathematical quantities as roots of coordinate wave
functions $\psi^{(0)}_k(0) = 0$ for $k = 1,3,4$, relative experimental
uncertainties should be attributed to $E_0$. This gives $\Delta E_0 =
1.6\times 10^{-15}\,{\rm eV}$ and $\Delta E_0 = 1.7\times
10^{-15}\,{\rm eV}$, respectively, which define the current
sensitivity of the qBounce experiments $\Delta E < 2\times
10^{-15}\,{\rm eV}$ \cite{Cronenberg2018}.  Below as an example, we
perform a numerical analysis of corrections, caused by interactions
violating Lorentz invariance, to transition frequencies of transitions
$|1\rangle \to |4\rangle$ between quantum gravitational states
$|1\rangle$ and $|4\rangle$ of unpolarized and polarized UCNs using
the current sensitivity of the qBounce experiments $\Delta E < 2\times
10^{-15}\,{\rm eV}$ \cite{Cronenberg2018}, i.e. $|\delta \nu_{p q}| <
\Delta E/2\pi$, where $\delta \nu_{pq}$ is a correction to the
transition frequency $\nu_{pq}$ of the transition $|q\rangle \to
|p\rangle$ between quantum gravitational states $|q\rangle$ and
$|p\rangle$ of unpolarized and polarized UCNs.

\subsection{\bf Corrections to binding energies of quantum
  gravitational states of unpolarized UCNs}

The problem of the calculation of corrections to the binding energies
of quantum gravitational states of unpolarized UCNs, induced by the
potential Eq.(\ref{eq:4}) violating Lorentz invariance, concerns the
stationary perturbation theory for degenerate quantum bound states
\cite{LL1965, Davydov1965}. Indeed, because of spin degrees of freedom
every quantum gravitational state of unpolarized UCNs is 2--fold
degenerate \cite{Davydov1965}.  In the zeroth approximation the
correct wave function of an unpolarized UCN in the $k$--quantum
gravitational state should be taken in the following form
\cite{LL1965}
\begin{eqnarray}\label{eq:6}
\Psi_k(z) = \psi^{(0)}_k(z)\,c_{\uparrow}\,\chi_{\uparrow} +
\psi^{(0)}_k(z)\,c_{\downarrow}\,\chi_{\downarrow},
\end{eqnarray}
where $\psi^{(0)}_k(z)$ is the coordinate wave function of UCNs in the
$k$--quantum gravitational state, $\chi_{\uparrow}$ and
$\chi_{\downarrow}$ are Pauli spinorial wave functions of the UCN in
the spin eigenstates {\it up} and {\it down}, respectively. Then, the
coefficients $c_{\uparrow}$ and $c_{\downarrow}$ are normalized by
$|c_{\uparrow}|^2 + |c_{\downarrow}|^2 = 1$ and define probabilities
to find the UCN in the $k$--quantum gravitational state with spin {\it
  up} and {\it down}, respectively. The coefficients $c_{\uparrow}$
and $c_{\downarrow}$ are determined in the first order approximation
to the binding energies \cite{LL1965, Davydov1965}.

For the calculation of the first order corrections to the binding
energies of quantum gravitational states of unpolarized UCNs we
rewrite the total Hamilton operator ${\rm H}$ of UCNs in the
gravitational field of the Earth with the potential $\Phi_{nLV}$ (see
Eq.(\ref{eq:5})) as follows
\begin{eqnarray}\label{eq:7}
{\rm H} = {\rm H}_0 + K_{\rm nLV} + Q_{\rm nLV \ell}\, S_{\ell},
\end{eqnarray}
where the operators $K_{\rm nLV}$ and $Q_{\rm nLV \ell}$ are given by
\begin{eqnarray}\label{eq:8}
&&K_{\rm nLV} = \big(- \bar{a}^n_j + m(\bar{c}^n_{0j} +
  \bar{c}^n_{j0}) + m \bar{e}^n_j\big) \frac{p_j}{m} - \big(2
  \bar{c}^n_{jk} + \bar{c}^n_{00}\delta_{jk}\big)\frac{p_j p_k}{2 m}+
  \frac{1}{2}\,(a_j \delta_{k\ell} - m e_j \delta_{k\ell})\,\frac{p_j
    p_k p_{\ell}}{m^3},\nonumber\\ &&Q_{\rm nLV \ell} = \big(- 2
  \bar{b}^n_{\ell} + 2 m \bar{d}^n_{\ell 0} - m \varepsilon_{\ell n m}
  \bar{g}^n_{n m 0} + \varepsilon_{\ell n m} \bar{H}^n_{n m}\big) +
  \Big(2 \bar{b}^n_0 \delta_{j\ell} - 2 m(\bar{d}^n_{\ell j} +
  \bar{d}^n_{00} \delta_{\ell j}) - m\, \varepsilon_{\ell n
    m}\big(\bar{g}^n_{m n j} + 2 \bar{g}^n_{m00}\delta_{j
    n}\big)\nonumber\\ \hspace{-0.15in}&& - 2 \varepsilon_{j \ell n}
  \bar{H}^n_{n 0}\Big)\frac{p_j}{m} + \Big(\big(2 \bar{d}^n_{0j} +
  \bar{d}^n_{j0} - \frac{1}{2}\, \varepsilon_{j m n} \bar{g}^n_{m n
    0}\big)\delta_{k\ell} + \frac{1}{2}\, \varepsilon_{\ell m n}
  \bar{g}^n_{mn0}\, \delta_{jk} - \varepsilon_{j \ell
    m}\,(\bar{g}^n_{m 0 k} + \bar{g}^n_{m k 0})\Big)\, \frac{p_j
    p_k}{m}\nonumber\\ \hspace{-0.15in}&& + \Big(\big( - \bar{b}^n_j -
  \frac{1}{2}\,\varepsilon_{j m n} \bar{H}^n_{mn}\big)\,\delta_{k\ell}
  + \bar{b}^n_{\ell}\Big)\,\frac{p_j p_k}{m^2} + \Big( \big(-
  \bar{b}^n_0\delta_{j \ell} + m \bar{d}^n_{\ell j} + \varepsilon_{j
    \ell n }\,\bar{H}^n_{n0}\big)\,\delta_{k m} + \big( - m
  \bar{d}^n_{jk} - m \frac{1}{2}\,\varepsilon_{k n p}\,\bar{g}^n_{n p
    j}\big)\,\delta_{m \ell}\Big)\nonumber\\ \hspace{-0.15in}&&\times
  \,\frac{p_j p_k p_m}{m^3}.
\end{eqnarray}
For the calculation of the first order correction to the energy level
of the $k$--quantum gravitational state of unpolarized UCNs we have to
solve the stationary Schr\"odinger--Pauli equation
\begin{eqnarray}\label{eq:9}
({\rm H} - E_k)\,\Psi_k(z) = 0,
\end{eqnarray}  
where $E_k = E^{(0)}_k + E^{(1)}_k$. Here $E^{(0)}_k = E_0 |\xi_k|$
and $E^{(1)}_k$ are the binding energy of the unperturbed $k$--quantum
gravitational state of unpolarized UCNs and the first order correction
to the energy level \cite{LL1965}, respectively. Following
\cite{LL1965, Davydov1965} Eq.(\ref{eq:9}) can be reduced to the
system of homogeneous algebraical equations for the coefficients
$c_{\uparrow}$ and $c_{\downarrow}$, which can have non--trivial
solutions if the determinant of this system is equal to zero
\begin{eqnarray}\label{eq:10}
 \left|\begin{array}{lcl} \langle \uparrow k |H|k \uparrow \rangle -
 E_k &\langle \uparrow k |H|k \downarrow \rangle\\ ~~~~\langle
 \downarrow k |H|k \uparrow \rangle & \langle \downarrow k |H|k
 \downarrow \rangle - E_k\\
\end{array}\right| = 0.
\end{eqnarray}
Such an equation is also called the {\it secular equation}
\cite{LL1965, Davydov1965}. Substituting the roots of the {\it secular
  equation} Eq.(\ref{eq:10}) into the system of algebraical equations
for the coefficients $c_{\uparrow}$ and $c_{\downarrow}$ and solving
it we determine the wave function of the $k$--quantum gravitational
state of an unpolarized UCN in the zeroth approximation \cite{LL1965}.

The matrix elements in Eq.(\ref{eq:10}) are defined by 
\begin{eqnarray}\label{eq:11}
\langle \uparrow k |H|k \uparrow \rangle &=& E^{(0)}_k + \langle
k|K_{\rm nLV}|k\rangle + \langle k|Q_{\rm nLV \ell}|k\rangle \langle
\uparrow |S_{\ell}|\uparrow \rangle,\nonumber\\ \langle \downarrow k
|H|k \downarrow \rangle &=& E^{(0)}_k + \langle k|K_{\rm nLV}|k\rangle
+ \langle k|Q_{\rm nLV \ell}|k\rangle \langle \downarrow
|S_{\ell}|\downarrow \rangle,\nonumber\\ \langle \uparrow  k |H|k 
\downarrow \rangle &=& \langle k|Q_{\rm nLV \ell}|k\rangle \langle
\uparrow |S_{\ell}|\downarrow \rangle \quad,\quad \langle \downarrow k
|H|k \uparrow \rangle = \langle k|Q_{\rm nLV \ell}|k\rangle \langle
\downarrow |S_{\ell}|\uparrow \rangle.
\end{eqnarray}
In order to simplify the solution of the {\it secular equation} we
 calculate the matrix elements of the neutron spin operator. For this
aim we may choose the axis of the neutron spin quantization along one
of the coordinate axes $(x,y,z)$. As a result, for the solution of the
{\it secular equation} we obtain the following expression
\begin{eqnarray}\label{eq:12}
E^{(\pm)}_k = E^{(0)}_k + \langle k|K_{\rm nLV}|k\rangle \pm
\frac{1}{2}\,\sqrt{ \big(\langle k|Q_{\rm nLV \ell}|k\rangle\big)^2}
\end{eqnarray}
with a summation over all components of the matrix element $\langle
k|Q_{\rm nLV \ell}|k\rangle$.  The second term in Eq.(\ref{eq:12})
defines the shift of the energy level, whereas the third one provides
a splitting of the energy level of the $k$--quantum gravitational
state of unpolarized UCNs into two levels separated by $\Delta E_k =
E^{(+)}_k - E^{(-)}_k = 2 E^{(1)}_k = \sqrt{ \big(\langle k|Q_{\rm nLV
    \ell}|k\rangle\big)^2}$. Such a splitting is fully caused by the
spin-dependent interaction violating Lorentz invariance in
Eq.(\ref{eq:4}). For the calculation of the matrix elements $\langle
k|K_{\rm nLV}|k\rangle$ and $\langle k|Q_{\rm nLV \ell}|k\rangle$ we
use the following integrals
\begin{eqnarray}\label{eq:13}
  \big\langle k|k\big\rangle &=& \int^{\infty}_0
  dz\,\psi^{(0)*}_k(z)\psi^{(0)}_k(z) = 1,\nonumber\\ \Big\langle
  k\Big|\frac{p_a p_b}{m^2}\Big|k\Big\rangle &=& \int^{\infty}_0
  dz\,\psi^{(0)*}_k(z)\frac{p_a p_b }{m^2}\,\psi^{(0)}_k(z) =
  \frac{2}{3}\,\frac{E^{(0)}_k}{m}\,\delta_{a z}\,\delta_{b z} =
  2.9\times 10^{-21}\,\frac{E^{(0)}_k}{E^{(0)}_4}\,\delta_{a
    z}\,\delta_{b z}, \nonumber\\ \Big\langle k\Big|\frac{p_a p_b
    p_c}{m^3}\Big|k\Big\rangle &=& \int^{\infty}_0
  dz\,\psi^{(0)*}_k(z)\frac{p_a p_b p_c}{m^3}\,\psi^{(0)}_k(z) =
  i\,\frac{g}{m}\,\delta_{a z}\,\delta_{b z}\,\delta_{c z} =
  i\,2.3\times 10^{-32}\,\delta_{a z}\,\delta_{b z}\,\delta_{c z},
\end{eqnarray}
which are obtained with the help of the relations derived by Albright
\cite{Albright1977}. One may also show that the non--vanishing value
of the last integral in Eq.(\ref{eq:13}) can be explained by a
non--vanishing first derivative of the wave functions of quantum
gravitational states of UCNs at the boundary $z = 0$, which is equal
to $d\psi^{(0)}_k(z)/dz\big|_{z = 0} = \ell^{-3/2} = \sqrt{2 m^2 g}$.

The numerical analysis shows that the contributions of the terms
proportional to $p_ap_bp_c/m^3$ can be neglected in comparison to the
contributions of other terms in the potential Eq.(\ref{eq:4}). As a
result, the matrix elements $\langle k|K_{\rm nLV}|k\rangle$ and
$\langle k|Q_{\rm nLV \ell}|k\rangle$ are equal to
\begin{eqnarray}\label{eq:14}
 \hspace{-0.3in} \langle k|K_{\rm nLV}|k\rangle &=& - \,\big(2
 m\,\bar{c}^n_{zz} + m\,
 \bar{c}^n_{00}\big)\,\frac{~E^{(0)}_k}{3m}\quad,\quad \langle
 k|Q_{\rm nLV x}|k\rangle = - 2 \tilde{b}^n_x + 2 (\bar{b}^n_x -
 \,m\,\bar{g}_{y0z})\,
 \frac{~E^{(0)}_k}{3m},\nonumber\\ \hspace{-0.3in} \, \langle k|Q_{\rm
   nLV y}|k\rangle &=& - 2 \tilde{b}^n_y + 2 (\bar{b}^n_y + m\,
 \bar{g}_{x0z})\, \frac{~E^{(0)}_k}{3m}\quad, \quad \langle k|Q_{\rm
   nLV z}|k\rangle = - 2 \tilde{b}^n_z + \,4\,
 \tilde{d}^n_z\,\frac{~E^{(0)}_k}{3 m}.
\end{eqnarray}
where we have used the notations $\tilde{b}^n_{\ell} =
\bar{b}^n_{\ell} - m \bar{d}^n_{\ell 0} + \frac{1}{2}\,m
\varepsilon_{\ell n m} \bar{g}^n_{n m 0} -
\frac{1}{2}\,\varepsilon_{\ell n m} \bar{H}^n_{n m}$ and
$\tilde{d}^n_z = m(\bar{d}^n_{0z} + \frac{1}{2}\,\bar{d}^n_{z0}) -
\frac{1}{2}\,\bar{H}^n_{xy}$ introduced in \cite{Kostelecky1999c,
  Kostelecky2011b}. Following the constraints on the parameters
$\tilde{b}_{\ell}$, i.e. $|\tilde{b}_{\ell}| < 7\times 10^{-30}\,{\rm
  GeV}$ \cite{Kostelecky2011b}, we may neglect the contributions of
$\tilde{b}_{\ell}$ to the matrix elements $\langle k|Q_{\rm nLV
  \ell}|k\rangle$. Using the constraint $|\bar{b}^n_{\ell} -
\frac{1}{2}\,\varepsilon_{\ell jk}\bar{H}^n_{jk}| < 10^{-28}\,{\rm
  GeV}$ for $\ell = x,y$ (see Ref.\cite{Kostelecky2011b}, Table 12,
p.35) and excluding occasional cancellation because of linear
independence of the parameters $\bar{b}^n_{\ell}$ and
$\bar{H}^n_{jk}$ we may also neglect the contributions of the
parameters $\bar{b}^n_{\ell}$ for $\ell = x,y$. As a result, we get
four transition frequencies of transitions $|q\rangle \to |p\rangle$
between quantum gravitational states $|q\rangle$ and $|p\rangle$ of
unpolarized UCNs
\begin{eqnarray}\label{eq:15}
\delta \nu_{pq} = - \,\big(2 m\bar{c}^n_{zz} + m
\bar{c}^n_{00}\big)\,\frac{E^{(0)}_p - E^{(0)}_q}{6\pi m} \pm \sqrt{(m
  \bar{g}_{x0z})^2 + ( m \bar{g}_{y0z})^2 + 4 \,
  \tilde{d}^2_z}\,\frac{E^{(0)}_p \pm E^{(0)}_q}{6\pi m}.
\end{eqnarray}
For the numerical analysis we shall use only the corrections where the
second term is proportional to $(E^{(0)}_p + E^{(0)}_q)$.

\subsection{\bf Corrections to binding energies of quantum
  gravitational states of polarized UCNs}

\begin{figure}
\includegraphics[height=0.26\textheight]{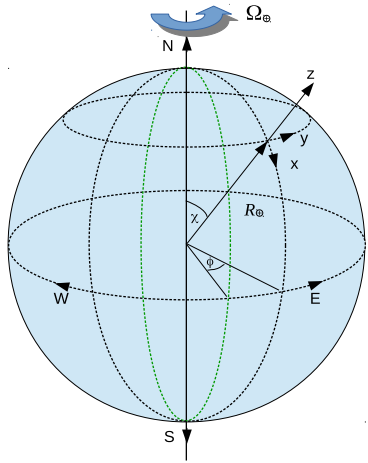}
  \caption{The position of the ILL laboratory of the qBounce
    experiments on the surface of the Earth.}
\label{fig:fig1}
\end{figure}

For the calculation of corrections to the binding energies of quantum
gravitational states of polarized UCNs we have to solve the
Schr\"odinger--Pauli equation Eq.(\ref{eq:9}), however with the wave
functions in the zeroth approximation taken in the following form
$\Psi_{k\sigma}(z) = \psi^{(0)}_k(z)\,\chi_{\sigma}$ with either
$\sigma = \uparrow$ or $\sigma = \downarrow$. Since in this case
quantum gravitational levels of UCNs are not degenerate with respect
to neutron spin--degrees of freedom, for the calculation of
corrections to the energy levels of quantum gravitational states of
UCNs we have to use the stationary perturbation theory for
non--degenerate bound states. Using Eq.(38.6) of Ref.\cite{LL1965} we
get the correction $E^{(1)}_{k\sigma}$ to the energy level of the
$k$--quantum gravitational state of polarized UCNs
\begin{eqnarray}\label{eq:16}
E^{(1)}_{k\sigma} = \int^{\infty}_0
dz\,\Psi^{\dagger}_{p\sigma'}(z)\Phi_{\rm nLV}\Psi_{p\sigma'}(z) =
\langle k|K_{\rm nLV}|k\rangle + \langle k|Q_{\rm nLV\ell}|k\rangle
\langle \sigma|S_{\ell}|\sigma\rangle.
\end{eqnarray}
The correction $\delta \nu^{(\ell)}_{p\sigma' q\sigma}$ to the
transition frequency $\nu_{p\sigma' q\sigma}$ of the transition
$|q\sigma\rangle \to |p\sigma'\rangle$ between quantum gravitational
states of polarized UCNs is equal to $\delta \nu^{(\ell)}_{p\sigma'
  q\sigma} = (E^{(1)}_{p\sigma'} - E^{(1)}_{q\sigma})/2\pi$, where
$\ell = x, y, z$ shows a direction of the quantization axis of the
neutron spin. For non--spin--flip transitions $|q\sigma\rangle \to
|p\sigma\rangle$ we get
\begin{eqnarray}\label{eq:17}
 \delta \nu^{(x)}_{p\sigma q\sigma} &=& \big(- (2 m \bar{c}^n_{zz} + m
 \bar{c}^n_{00}) - 2m\,\bar{g}^n_{y0z}\,\langle \sigma|
 S_x|\sigma\rangle\big)\, \frac{E^{(0)}_p - E^{(0)}_q}{6\pi m} =
 \big(- (2 m \bar{c}^n_{zz} + m \bar{c}^n_{00}) \mp
 m\,\bar{g}^n_{y0z}\big)\, \frac{E^{(0)}_p - E^{(0)}_q}{6\pi
   m},\nonumber\\ \delta \nu^{(y)}_{p\sigma q\sigma} &=& \big(- (2 m
 \bar{c}^n_{zz} + m \bar{c}^n_{00}) + 2m\,\bar{g}^n_{x0z}\,\langle
 \sigma| S_y|\sigma\rangle\big)\, \frac{E^{(0)}_p - E^{(0)}_q}{6\pi m}
 = \big(- (2 m \bar{c}^n_{zz} + m \bar{c}^n_{00}) \pm
 m\,\bar{g}^n_{x0z}\big)\, \frac{E^{(0)}_p - E^{(0)}_q}{6\pi
   m},\nonumber\\ \delta \nu^{(z)}_{p\sigma q\sigma} &=& \Big(- (2 m
 \bar{c}^n_{zz} + m \bar{c}^n_{00}) - 4\,\tilde{d}^n_z\,\langle
 \sigma| S_z|\sigma\rangle\Big)\, \frac{E^{(0)}_p - E^{(0)}_q}{6\pi m}
 = \big(- (2 m \bar{c}^n_{zz} + m \bar{c}^n_{00}) \mp
 2\,\tilde{d}^n_z\big)\, \frac{E^{(0)}_p - E^{(0)}_q}{6\pi m},
\end{eqnarray}
where $\langle \sigma| S_{\ell}|\sigma \rangle =
\chi^{\dagger}_{\sigma}S_{\ell}\chi_{\sigma} = \pm \frac{1}{2}$ is an
averaged value of the neutron spin operator $S_{\ell}$ for $\ell =
x,y,z$ with the quantization axis of the neutron spin directed along
$x$--, $y$-- and $z$--axis, respectively, in the standard laboratory
frame (see Fig.\,\ref{fig:fig1}). In turn, for spin--flip transitions
$|q\sigma\rangle \to |p\sigma'\rangle$ with $(\sigma = \uparrow,
\sigma' = \downarrow)$ or $(\sigma = \downarrow, \sigma' = \uparrow)$,
respectively, we obtain
\begin{eqnarray}\label{eq:18}
 \delta \nu^{(x)}_{p\sigma' q\sigma} = \mp m\,\bar{g}^n_{y0z}\,
 \frac{E^{(0)}_p + E^{(0)}_q}{6\pi m}\;,\; \delta \nu^{(y)}_{p\sigma'
   q\sigma} = \pm m\,\bar{g}^n_{x0z}\, \frac{E^{(0)}_p +
   E^{(0)}_q}{6\pi m}\;,\; \delta \nu^{(z)}_{p\sigma' q\sigma} = \mp
 2\,\tilde{d}^n_z\, \frac{E^{(0)}_p + E^{(0)}_q}{6\pi m}.
\end{eqnarray}
For the derivation of Eq.(\ref{eq:17}) and Eq.(\ref{eq:18}) we
have used the matrix elements in Eq.(\ref{eq:14}) taken in the
approximation discussed below Eq.(\ref{eq:14}). We would like to
mention that, of course, the measurement of transition frequencies of
non--spin--flip and spin--flip transitions between quantum
gravitational states of polarized UCNs is a nearest future for the
qBounce experiments.

\subsection{Numerical analysis of parameters of Lorentz invariance
  violation from transition frequencies in the standard laboratory
  frame}

For the numerical analysis of parameters of Lorentz invariance
violation we use the transitions $|1\rangle \to |4\rangle$ between
quantum gravitational states $|1\rangle$ and $|4\rangle$ for
unpolarized and polarized UCNs. From the spin--flip transitions we get
\begin{eqnarray}\label{eq:19}
 |\bar{g}^n_{y0z}| < \frac{3 \Delta E}{E^{(0)}_4 + E^{(0)}_1} =
 1.1\times 10^{-3}\,,\,|\bar{g}^n_{x0z}| < \frac{3 \Delta E}{E^{(0)}_4
   + E^{(0)}_1} = 1.1\times 10^{-3}\,,\, |\tilde{d}^n_z| <
 \frac{3}{2}\,\frac{\Delta E m}{E^{(0)}_4 + E^{(0)}_1} = 5.1\times
 10^{-4} {\rm GeV}.
\end{eqnarray}
Using these constraints we may estimate the value of the parameter $(2
\bar{c}^n_{zz} + \bar{c}^n_{00})$. From the analysis of the
corrections to the transition frequencies of transitions of
unpolarized UCNs Eq.(\ref{eq:15}) and of non--spin--flip transitions
of polarized UCNs Eq.(\ref{eq:17}) we get
\begin{eqnarray}\label{eq:20}
| 2 \bar{c}^n_{zz} + \bar{c}^n_{00}| < \frac{6 \Delta
  E\,E^{(0)}_4}{E^{(0)2}_4 - E^{(0)2}_1} = 3.4\times 10^{-3}.
\end{eqnarray}
Thus, measurements of transition frequencies of transitions between
quantum gravitational states of unpolarized and polarized UCNs allow
to place some new constraints on the parameters of Lorentz invariance
violation in comparison to the results adduced in
Ref.\cite{Kostelecky2011b}.

\section{Parameters of violation of Lorentz  invariance in the
  canonical Sun--centered frame}
\label{sec:sun}

Values of parameters of Lorentz invariance violation should in
principle depend on an inertial frame, where they are measured. In the
ground--based laboratory on the surface of the Earth such as the ILL,
i.e.in the standard laboratory frame with coordinates $(t, x, y, z)$
(see Fig.\,\ref{fig:fig1}), parameters of violation of Lorentz
invariance should vary in time with a period ${\cal T}_{\oplus} = 23\,
{\rm hr}\, 56\, {\rm min}\, 4.09\,{\rm s}$ determined by the Earth's
sidereal angular frequency $\Omega_{\oplus} = 2\pi/{\cal T}_{\oplus} =
7.2921159 \times 10^{-5}\,{\rm rad/s}$. It is obvious that because of
rotation, yielding distinguishable inertial frames in a ground--based
laboratory on the surface of the Earth, the standard
  laboratory frame is not appropriate for definition of the values of
parameters of Lorentz invariance violation. In contrast, the frame
centered on the Sun, i.e. the canonical Sun--centered frame, remains
unchanged approximately inertial frame over thousands of years
\cite{Kostelecky1999c}--\cite{Kostelecky2016}. Thus, following
\cite{Kostelecky1999c}--\cite{Kostelecky2016} we define parameters of
violation of Lorentz invariance in Eq.(\ref{eq:4}) in terms of the
parameters of violation of Lorentz invariance in the canonical
Sun--centered frame with coordinates $(T, X, Y, Z)$ (see also
\cite{Smart1977}), where $T$ is the celestial equatorial time An
important point of the expression of parameters of violation of
Lorentz invariance in the laboratory frame in terms of parameters in
the canonical Sun--centered one we have to relate a local laboratory
time $t$ to a time $T$ in the canonical Sun--centered frame, where $T$
is the celestial equatorial time \cite{Kostelecky2002b}. Such a
problem has been discussed in details in
\cite{Kostelecky1999c}--\cite{Kostelecky2016}.  According to
\cite{Kostelecky1999c}--\cite{Kostelecky2016}, it is useful to match
$t$ with the local sidereal time $T_{\oplus}$, which is measured in
the canonical Sun--centered frame from one of the times when the $y$
axis lies along the $Y$ axis. The time $T{\oplus}$ is related to the
celestial equatorial time $T$ by the relation \cite{Kostelecky2016}
\begin{eqnarray}\label{eq:21}
T_{\oplus} = T - T_0\quad,\quad T_0 = \frac{66.25^0 -
  \phi}{360^0}\,(23.934\,{\rm hr}),
\end{eqnarray}
where $\phi$ is a longitude of the laboratory measured in degrees. As
the longitude of the ILL laboratory is $\phi = 5.71667^0$, we get
$T_{\oplus} = T - 4\,{\rm hr}\,01\,{\rm min}\,30.44\,{\rm
  s}$. According to \cite{Kostelecky1999c}--\cite{Kostelecky2016},
with reasonable approximation that the orbit of the Earth is circular,
the transition from the canonical Sun--centered frame with coordinates
$(X, Y, Z)$ to the laboratory frame with coordinates $(x, y, z)$ is
given by the matrix dependent on the sidereal time $T_{\oplus}$
\cite{Kostelecky1999c}--\cite{Kostelecky2016}
\begin{eqnarray}\label{eq:22}
R_{jJ}(T_{\oplus}) = \left(\begin{array}{llcl} \cos \chi \cos
  \Omega_{\oplus}T_{\oplus} & \cos\chi \sin\Omega_{\oplus} T_{\oplus}
  & - \sin\chi\\ - \sin\Omega_{\oplus} T_{\oplus} & \cos
  \Omega_{\oplus} T_{\oplus} & 0\\ \sin \chi \cos \Omega_{\oplus}
  T_{\oplus} & \sin \chi \sin \Omega_{\oplus}T_{\oplus} & \cos\chi \\
\end{array}\right),
\end{eqnarray}
where $j = x,y,z$ and $J = X, Y, Z$ denote the indices in the
laboratory and canonical Sun--centered frames, respectively. The
matrix $R(T_{\oplus})$ in Eq.(\ref{eq:22}) obeys the constraint
$R(T_{\oplus})R^T(T_{\oplus}) = R^T(T_{\oplus}) R(T_{\oplus}) = 1$,
where $T$ is a transposition.  Now we may express parameters violating
Lorentz invariance in the laboratory frame in terms of the parameters
in the canonical Sun--centered frame. This concerns only parameters
entering in Eq.(\ref{eq:17}), Eq.(\ref{eq:18}) and Eq.(\ref{eq:20}).
We get
\begin{eqnarray}\label{eq:23}
\bar{c}^n_{zz} &=& R_{zA}(T_{\oplus})R_{zB}(T_{\oplus})\bar{c}^n_{AB}
= \frac{1}{2}\,\big[\sin^2\chi\big(\bar{c}^n_{XX} +
  \bar{c}^n_{YY}\big) + 2 \cos^2\chi\,\bar{c}^n_{ZZ}\big] + \sin\chi
\cos\chi\,[\big(\bar{c}^n_{XZ} + \bar{c}^n_{ZX}) \cos \Omega_{\oplus}
  T_{\oplus}\nonumber\\ &+& \big(\bar{c}^n_{YZ} + \bar{c}^n_{ZY}\big)
  \sin \Omega_{\oplus} T_{\oplus}\big] +
\frac{1}{2}\,\sin^2\chi\big(\bar{c}^n_{XX} - \bar{c}^n_{YY}\big)\,\cos
2\Omega_{\oplus}T_{\oplus},\nonumber\\
\bar{d}^n_{0z} &=& R_{zJ}(T_{\oplus}) \bar{d}^n_{TJ} =
\sin\chi\big(\bar{d}^n_{TX}\cos \Omega_{\oplus}T_{\oplus} +
\bar{d}^n_{TY} \sin \Omega_{\oplus}T_{\oplus}\big) +
\cos\chi\,\bar{d}^n_{TZ},\nonumber\\ \bar{d}^n_{z0} &=&
R_{zJ}(T_{\oplus}) \bar{d}^n_{JT} = \sin\chi\big(\bar{d}^n_{XT} \cos
\Omega_{\oplus}T_{\oplus} + \bar{d}^n_{YT} \sin \Omega_{\oplus}
T_{\oplus}\big) + \cos\chi\,\bar{d}^n_{ZT},\nonumber\\ \bar{g}^n_{y 0
  z} &=&R_{y A}(T_{\oplus})R_{z B}(T_{\oplus}) \bar{g}^n_{ATB} =
\frac{1}{2}\,\sin\chi\,\big(\bar{g}^n_{YTX} - \bar{g}^n_{XTY}\big) +
\cos\chi\,\big(\bar{g}^n_{YTZ}\,\cos\Omega_{\oplus} T_{\oplus} -
\bar{g}^n_{XTZ}\,\sin\Omega_{\oplus} T_{\oplus}\big) \nonumber\\ &+&
\frac{1}{2}\,\sin\chi\,\big[\big(\bar{g}^n_{XTY} +
  \bar{g}^n_{YTX}\big)\,\cos 2\Omega_{\oplus} T_{\oplus} -
  \big(\bar{g}_{XTX} - \bar{g}_{YTY}\big)\,\sin 2\Omega_{\oplus}
  T_{\oplus}\big], \nonumber\\ \bar{g}^n_{yz0}
&=&R_{yA}(T_{\oplus})R_{z B}(T_{\oplus}) \bar{g}^n_{ABT} = -
\sin\chi\, \bar{g}^n_{XYT} + \cos\chi\,\big(\bar{g}^n_{YZT}
\,\cos\Omega_{\oplus} T_{\oplus} - \bar{g}^n_{XZT}\,\sin
\Omega_{\oplus} T_{\oplus}\big), \nonumber\\ \bar{g}^n_{x 0 z} &=&
R_{x A}(T_{\oplus})R_{z B}(T_{\oplus}) \bar{g}^n_{ATB} = \frac{1}{2}
\sin\chi \cos\chi\,\big(\bar{g}^n_{XTX} + \bar{g}^n_{YTY} - 2
\bar{g}^n_{ZTZ} \big) - \big(\sin^2\chi \bar{g}^n_{ZTX} -
\cos^2\chi\,\bar{g}^n_{XTZ}\big)\nonumber\\ &&\times\,\cos
\Omega_{\oplus} T_{\oplus} - \big(\sin^2\chi \bar{g}^n_{ZTY} -
\cos^2\chi \bar{g}^n_{YTZ}\big) \sin \Omega_{\oplus} T_{\oplus} +
\frac{1}{2}\,\sin\chi \cos\chi\,\big[\big( \bar{g}^n_{XTX} -
  \bar{g}^n_{YTY}\big)\,\cos 2\Omega_{\oplus}
  T_{\oplus}\nonumber\\ &+& \big(\bar{g}^n_{XTY} +
  \bar{g}^n_{YTX}\big) \sin 2\Omega_{\oplus} T_{\oplus}\big],
\nonumber\\ \bar{g}^n_{xz0} &=& R_{xA}(T_{\oplus})R_{z B}(T_{\oplus})
\bar{g}^n_{ABT}= - \bar{g}^n_{ZXT}\,\cos \Omega_{\oplus} T_{\oplus} -
\bar{g}^n_{ZYT}\,\sin \Omega_{\oplus}
T_{\oplus},\nonumber\\ \bar{b}^n_j &=&
R_{jJ}(T_{\oplus})\bar{b}^n_J,\nonumber\\ \bar{d}^n_{j0}
&=&R_{jJ}(t)\bar{d}^n_{JT},\nonumber\\ \varepsilon_{jk\ell}
\bar{g}^n_{kl0} &=& R_{jJ}(T_{\oplus})\varepsilon_{JKL}
\bar{g}^n_{KLT},\nonumber\\ \varepsilon_{jk\ell} \bar{H}^n_{k\ell} &=&
R_{jJ}(T_{\oplus}) \varepsilon_{JKL} \bar{H}^n_{KL}.
\end{eqnarray}
As has been pointed out in \cite{Kostelecky2002b} the local sidereal
time $T_{\oplus}$ should be chosen conveniently for every
experiment. This can be also done defining the local sidereal time
$T_{\oplus}$ in terms of the local laboratory time $t$ as follows
$T_{\oplus} = t - t_0$, where $t_0$ can be determined for every run of
qBounce experiments \cite{LSTime}.

The transition frequencies for spin--flip and non--spin--flip
transitions averaged over time impose the following upper bounds on
parameters of violation of Lorentz invariance
\begin{eqnarray}\label{eq:24}
|\bar{g}^n_{XTY} - \bar{g}^n_{YTX}| &&< \frac{2}{\sin\chi}\times
1.1\times 10^{-3} = 3.1\times 10^{-3},\nonumber\\
|\tilde{g}^n_Q| &&<
\frac{2m}{\sin\chi\,\cos\chi}\times 1.1\times 10^{-3} = 4.1\times
10^{-3}\,{\rm GeV},\nonumber\\
|\tilde{d}^n_Z| &&< \frac{1}{2\,\cos\chi}\times
5.1\times 10^{-4} \,{\rm GeV} = 3.6\times 10^{-4}\,{\rm GeV},
\end{eqnarray}
where we have used the notation $\tilde{g}^n_Q = m \big(\bar{g}^n_{XTX}
+ \bar{g}^n_{YTY} - 2 \bar{g}^n_{ZTZ}\big)$ \cite{Kostelecky1999c},
and
\begin{eqnarray}\label{eq:25}
\big|(1 + \sin^2\chi)\,\tilde{c}^n_Q + 5\,m\,\bar{c}^n_{ZZ}\big| < 3.4
 \times 10^{-3}m,
\end{eqnarray}
where we have used the notation $\tilde{c}^n_Q = m (\bar{c}^n_{XX} +
\bar{c}^n_{YY} - 2 \bar{c}^n_{ZZ})$ \cite{Kostelecky1999c} and the
traceless of $\bar{c}^n_{\mu\nu}$, i.e. $\bar{c}^n_{TT} =
\bar{c}^n_{XX} + \bar{c}^n_{YY} + \bar{c}^n_{ZZ}$. Since
$\tilde{c}^n_Q = (- 1.8 \pm 2.2)\times 10^{-14}\,{\rm GeV}$ (see
Ref.\cite{Kostelecky2011b} (see p.36, Table D12)), we may neglect the
contribution of $\tilde{c}^n_Q$ and place a new constraint
$|\bar{c}^n_{ZZ}| < 6.8\times 10^{-4}$.  We would like to emphasize
that so far the parameter $|\bar{c}^n_{ZZ}|$ was not yet estimated
(see Ref.\,\cite{Kostelecky2011b}, p.36, Table D12). Then, using
$\tilde{c}^n_Q = (- 1.8 \pm 2.2)\times 10^{-14}\,{\rm GeV}$ and
$\bar{c}^n_{XX} - \bar{c}^n_{YY} = (1.4 \pm 1.7)\times 10^{-29}$ (see
Ref.\,\cite{Kostelecky2011b}, p.36, Table D12), and $|\bar{c}^n_{ZZ}|
< 6.8\times 10^{-4}$ one may assume that $|\bar{c}^n_{XX}| =
|\bar{c}^n_{YY}| < 6.8 \times 10^{-4}$. Using the property
$\eta^{\mu\nu} \bar{c}^n_{\mu\nu} = 0$ of the tensor
$\bar{c}^n_{\mu\nu}$ we get $|\bar{c}^n_{TT}| = 3 |\bar{c}^n_{ZZ}| <
2.0 \times 10^{-3}$. Of course, such an estimate we can make also as
follows. Expressing the parameter $m \bar{c}^n_{ZZ}$ in terms of
$\bar{c}^n_Q$ and $m \bar{c}^n_{TT}$, we get $m\bar{c}^n_{ZZ} = (m
\bar{c}^n_{TT} - \bar{c}^n_Q)/3$ and transcribe Eq.(\ref{eq:25})
into the form
\begin{eqnarray}\label{eq:26}
 \Big|\Big(\sin^2\chi - \frac{2}{3}\Big)\,\bar{c}^n_Q +
 \frac{5}{3}\,m\,\bar{c}^n_{TT}\Big| < 3.4 \times 10^{-3}\,m.
\end{eqnarray}
Because of the experimental data $\tilde{c}^n_Q = (- 1.8 \pm
2.2)\times 10^{-14}\,{\rm GeV}$ giving $\bar{c}^n_{ZZ} =
(\bar{c}^n_{XX} + \bar{c}^n_{YY})/2$, the experimental data
$\bar{c}^n_{XX} - \bar{c}^n_{YY} = (1.4 \pm 1.7)\times 10^{-29}$
giving $\bar{c}^n_{XX} = \bar{c}^n_{YY} = \bar{c}^n_{ZZ}$ and the
relation $\bar{c}^n_{ZZ} = \bar{c}^n_{TT}/3$, we arrive at the
constraint $|\bar{c}^n_{ZZ}| < 6.8\times 10^{-4}$.

 In Table I we have collected the obtained results in the form
 accepted in \cite{Kostelecky2011b}. Of course, all of these estimates
 should be treated as a theoretical basis for future qBounce
 experiments of searches for effects of interactions violating Lorentz
 invariance in the neutron sector of the SME.
\begin{table}[h]
\begin{tabular}{|c|c|}
  \hline Combination & Result \\\hline $|\bar{c}^n_{XX}|$ & $<
  6.8\times 10^{-4}$ \\ \hline $|\bar{c}^n_{YY}|$ & $< 6.8\times
  10^{-4}$ \\ \hline $|\bar{c}^n_{ZZ}|$ & $< 6.8\times
  10^{-4}$\\\hline $|\bar{c}^n_{TT}|$ & $< 2.0\times 10^{-3}$
  \\ \hline $|\bar{g}^n_{XTY} - \bar{g}^n_{YTX}|$ & $< 3.1 \times
  10^{-3}$\\\hline $|\tilde{g}_Q|$ & $< 4.1 \times 10^{-3}\,{\rm
    GeV}$\\ \hline $\big|\tilde{d}^n_Z \big|$ & $ < 3.6 \times
  10^{-4}\,{\rm GeV}$\\\hline
\end{tabular} 
\caption{Neutron sector. A theoretical basis for an experimental
  analysis of upper bounds of parameters of Lorentz invariance
  violation in the neutron sector of the SME obtained by using the
  current experimental sensitivity $\Delta E < 2 \times 10^{-15}\,{\rm
    eV}$ of qBounce experiments \cite{Cronenberg2018}. }
\end{table}

\section{Evolution of spin operator of UCNs}
\label{sec:spin}

A time evolution of the spin operator $\vec{S}$ of UCNs is described
by Heisenberg's equation of motion \cite{LL1965}
\begin{eqnarray}\label{eq:27}
\frac{d\vec{S}}{dt} = \frac{\partial \vec{S}}{\partial t} + i\,[{\rm
    H}, \vec{S}\,] = \frac{\partial \vec{S}}{\partial t} +
i\,[\Phi_{\rm nLV},\vec{S}\,].
\end{eqnarray}
Since the spin operator $\vec{S}$ does not depend explicitly on time,
the partial derivative in Eq.(\ref{eq:18}) is equal to zero. Then,
using the effective non--relativistic potential Eq.(\ref{eq:4}) and
the commutation relation $[S_a, S_b] = i\varepsilon_{abc}S_c$ we get
\begin{eqnarray}\label{eq:28}
\frac{d\vec{S}}{dt} = \vec{\Omega}_{\rm nLV}\times \vec{S},
\end{eqnarray}
where $\vec{\Omega}_{\rm nLV}$ is the angular velocity operator, induced by
violation of Lorentz  invariance. It is equal to
\begin{eqnarray}\label{eq:29}
(\vec{\Omega}_{\rm nLV})_a &=& - 2 \tilde{b}^n_a + \Big(2 b^n_0
  \delta_{j a} - 2 m(d^n_{aj} + d^n_{00} \delta_{aj}) - m\,
  \varepsilon_{a \ell m}\big(g^n_{m\ell j} + 2
  g_{m00}\delta_{j\ell}\big) - 2 \varepsilon_{ja\ell} H^n_{\ell
    0}\Big)\frac{p_j}{m}\nonumber\\ &+& \Big\{\Big(- b^n_j + 2
  \tilde{d}^n_j - \frac{1}{2}\,m\,\varepsilon_{j m n}\,g^n_{m n
    0}\Big)\delta_{k a} + b^n_a + \frac{1}{2}\,m\, \varepsilon_{a m n}
  g^n_{mn0} \delta_{jk} - m\, \varepsilon_{j a m}\,(g^n_{m 0 k} +
  g^n_{m k 0})\Big\}\frac{p_j p_k}{m^2}.
\end{eqnarray}
Thus, Eq.(\ref{eq:28}) with the angular velocity operator
Eq.(\ref{eq:29}) defines an spin evolution of UCNs caused by
interactions violating Lorentz invariance. A dependence of the
parameters violating Lorentz invariance on time and the Earth's
sidereal frequency leads to fluctuations of the angular velocity $\vec
{\Omega}_{\rm nLV}$ with a period ${\cal T}_{\oplus} = 23\,{\rm
  hr}\,56\,{\rm min}\, 4.09\,{\rm s}$ of the Earth's rotation.

\section{Discussion}
\label{sec:discussion}

In this paper we have proposed to extract constraints on the
parameters of Lorentz invariance violation from measurements of
transition frequencies of transitions between quantum gravitational
states of unpolarized and polarized UCNs by the qBounce Collaboration
\cite{Cronenberg2018, Ivanov2013}. Neglecting contributions of
parameters of violation of Lorentz invariance in the gravitational
sector of the SME, and using the effective non--relativistic potential
for interactions violating Lorentz invariance in the neutron sector of
the SME, derived by Kosteleck\'y and Lane \cite{Kostelecky1999b}, we
have calculated the corrections, caused by interactions violating
Lorentz invariance, to transition frequencies of transitions between
quantum gravitational states of unpolarized and polarized UCNs. We
have carried out the calculations in \textcolor{red}{the standard
  laboratory frame} relative to the location of the ILL laboratory on
the surface of the Earth.

For the extraction of constraints on parameters of violation of
Lorentz invariance from these corrections it is important to define
parameters of violation of Lorentz invariance relative to an inertial
frame. Because of rotation of the Earth all terrestrial laboratories
are non--inertial and parameters of violation of Lorentz invariance,
analyzed in any terrestrial laboratory, should be expressed in terms
of parameters of violation of Lorentz invariance in an inertial frame,
for example, in the canonical Sun--centered frame, which remains
unchanged approximately inertial frame over thousands of years
\cite{Kostelecky1999c}--\cite{Kostelecky2016}.

Having expressed parameters of violation of Lorentz invariance, which
are responsible for corrections to the transition frequencies of
transitions between quantum gravitational states of unpolarized and
polarized UCNs in the ILL laboratory, in terms of parameters of
violation of Lorentz invariance defined in the canonical Sun--centered
frame, we have placed some new constraints (see Table I) in comparison
to the results represented in Ref.\cite{Kostelecky2011b} (see
Ref.\cite{Kostelecky2011b}, Table D12). In Table I we have adduced the
constraints on parameters of violation of Lorentz invariance using the
current sensitivity $\Delta E < 2\times 10^{-15}\,{\rm eV}$ of the
qBounce experiments \cite{Cronenberg2018}, which is closely related to
experimental uncertainties of the experimental data
\cite{Cronenberg2018} (see also a discussion below
Eq.(\ref{eq:16})). The results represented in Table I may serve as a
theoretical basis for experimental searches of effects of violation of
Lorentz invariance in the neutron sector of the SME in qBounce
experiments.

Thus, following the results given in Table I one may assert that even
for the current sensitivity $\Delta E < 2\times 10^{-15}\,{\rm eV}$ of
the qBounce experiments \cite{Cronenberg2018} the analysis of
transition frequencies of transitions between quantum gravitational
states of unpolarized and polarized UCNs may allow to place new
constraints on parameters of violation of Lorentz invariance in the
neutron sector of the SME. The qBounce experiments on an analysis of
contributions of interactions violating Lorentz invariance will be
carried out at Instrument PF2/UCN at ILL. A possible improvement of
experimental sensitivity of the qBounce experiments up to $\Delta E <
10^{-17}\,{\rm eV}$ in the nearest future and finally to reach a
sensitivity $\Delta E < 10^{-21}\,{\rm eV}$ \cite{Abele2010} should
allow to improve substantially upper bounds of parameters in Table
I. In our forthcoming publication we are planning to analyze
contributions of interactions violating Lorentz invariance from the
gravitational sector of the SME \cite{Kostelecky2006,
  Kostelecky2011a}.

\section{Acknowledgements}

We are  grateful to Alan Kosteleck\'y for fruitful discussions and
comments. The work of A. N. Ivanov was supported by the Austrian
``Fonds zur F\"orderung der Wissenschaftlichen Forschung'' (FWF) under
the contracts P31702-N27 and P26636-N20, and ``Deutsche
F\"orderungsgemeinschaft'' (DFG) AB 128/5-2. The work of M. Wellenzohn
was supported by the MA 23 (FH-Call 16) under the project ``Photonik -
Stiftungsprofessur f\"ur Lehre''.

\end{document}